\pdfoutput=1
\documentclass[sigconf]{acmart}

\AtBeginDocument{%
  \providecommand\BibTeX{{%
    \normalfont B\kern-0.5em{\scshape i\kern-0.25em b}\kern-0.8em\TeX}}}

\copyrightyear{2021}
\acmYear{2021}
\setcopyright{iw3c2w3}
\acmConference[WWW '21]{Proceedings of the Web Conference 2021}{April 19--23, 2021}{Ljubljana, Slovenia}
\acmBooktitle{Proceedings of the Web Conference 2021 (WWW '21), April 19--23, 2021,
Ljubljana, Slovenia}

\acmPrice{}
\acmDOI{10.1145/3442381.3449998}
\acmISBN{978-1-4503-8312-7/21/04}

\begin{document}

\title{STAN: Spatio-Temporal Attention Network 
for Next Location Recommendation}


\author[Yingtao Luo, Qiang Liu, Zhaocheng Liu]{Yingtao Luo$^{1}$, Qiang Liu$^{2,3,}$*, Zhaocheng Liu$^{4}$}

\makeatletter
\def\authornotetext#1{
 \g@addto@macro\@authornotes{%
 \stepcounter{footnote}\footnotetext{#1}}%
}
\makeatother

\authornotetext{Corresponding author.}

\affiliation{%
 \institution{$^1$University of Washington}
 \institution{$^2$Center for Research on Intelligent Perception and Computing, Institute of Automation, Chinese Academy of Sciences}
 \institution{$^3$School of Artificial Intelligence, University of Chinese Academy of Sciences}
 \institution{$^4$Renmin University of China}
 \country{}
}

\email{yl3851@uw.edu, qiang.liu@nlpr.ia.ac.cn, lio.h.zen@gmail.com}

\def\authors{Yingtao Luo, Qiang Liu, and Zhaocheng Liu}


\begin{abstract}
The next location recommendation is at the core of various location-based applications. Current state-of-the-art models have attempted to solve spatial sparsity with hierarchical gridding and model temporal relation with explicit time intervals, while some vital questions remain unsolved. Non-adjacent locations and non-consecutive visits provide non-trivial correlations for understanding a user's behavior but were rarely considered. To aggregate all relevant visits from user trajectory and recall the most plausible candidates from weighted representations, here we propose a Spatio-Temporal Attention Network (STAN) for location recommendation. STAN explicitly exploits relative spatiotemporal information of all the check-ins with self-attention layers along the trajectory. This improvement allows a point-to-point interaction between non-adjacent locations and non-consecutive check-ins with explicit spatio-temporal effect. STAN uses a bi-layer attention architecture that firstly aggregates spatiotemporal correlation within user trajectory and then recalls the target with consideration of personalized item frequency (PIF). By visualization, we show that STAN is in line with the above intuition. Experimental results unequivocally show that our model outperforms the existing state-of-the-art methods by 9-17\%.
\end{abstract}



\maketitle

\section{Introduction}
Next Point-of-Interest (POI) recommendation raises intensive studies in recent years owing to the growth of location-based services such as Yelp, Foursquare and Uber. The large volume of historical check-in data gives service providers invaluable information to understand user preferences on next movements, as the historical trajectories reveal the user's behavioral pattern in making every decision. Meanwhile, such a system can also provide users with the convenience to decide where to go and how to plan the day, based on previous visits as well as current status \cite{feng2015personalized, ExploringTemporalEffects, Next, Contextual, liu2014exploiting}.

Previous approaches have extensively studied various aspects and proposed many models to make a personalized recommendation. Early models mainly focus on sequential transitions, such as Markov chains \cite{rendle2010factorization}. Later on, recurrent neural networks (RNNs) with memory mechanism improved recommendation precision, inspiring following works \cite{hidasi2015session, zhu2017next, LSTPM, DeepMove} to propose RNN variants to better extract the long periodic and short sequential features of user trajectories. Besides sequential regularities, researchers have exploited temporal and spatial relation to assist sequential recommendation \cite{STRNN}. The recent state-of-the-art models fed time intervals and/or spatial distances between two consecutive visits to explicitly represent the effect of the spatiotemporal gap between each movement. Prior works have also addressed the sparsity problem of spatiotemporal information by discretely denoting time in hours and partitioning spatial areas by hierarchical grids \cite{yang2020location, ASPPA, lian2020geography}. Besides, they modified neural architectures \cite{STGN, Serm, Caser} or stacked extra modules \cite{ARNN, LSTPM, chen2020context} to integrate these additional information. 

\begin{figure}[t]
\centering
\includegraphics[width=0.9\columnwidth]{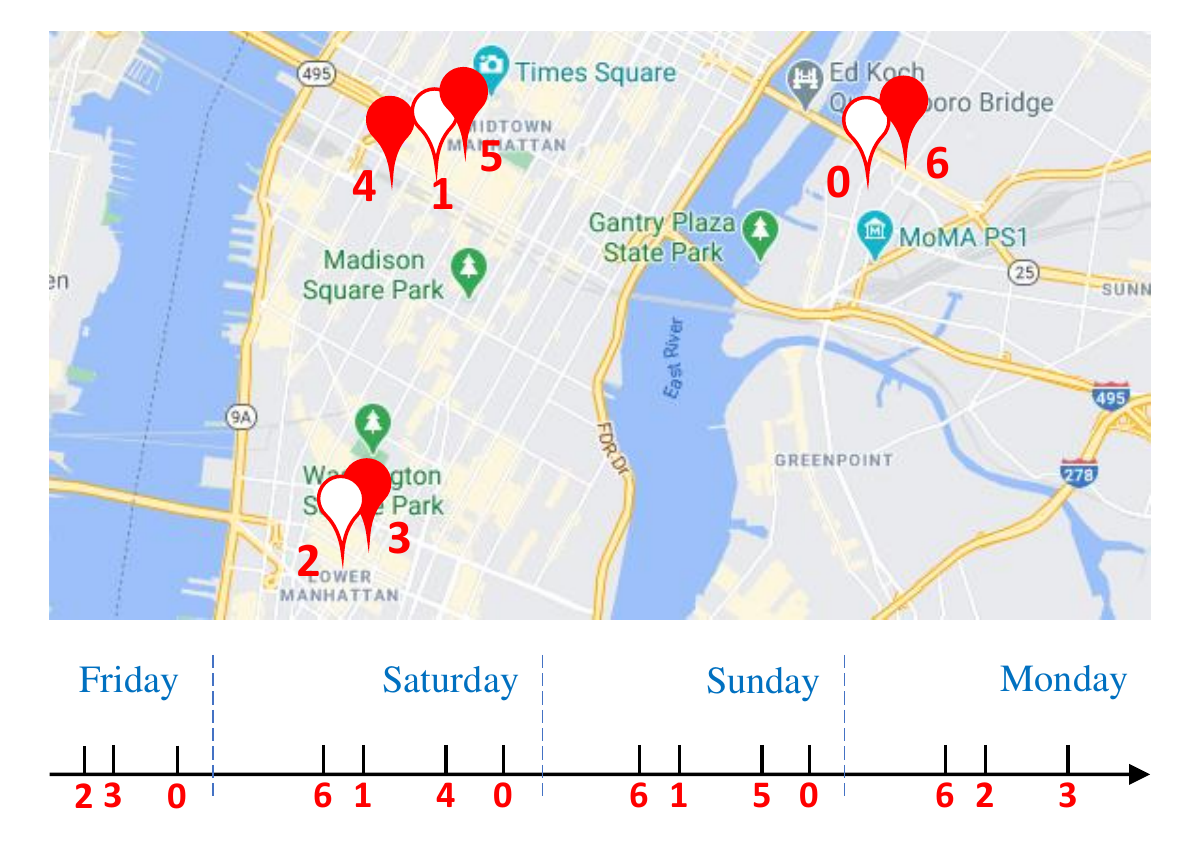}
\caption{A trajectory example showing the relation between non-consecutive visits and non-adjacent locations. The map shows the spatial distribution of visited locations, which are named by figures from 0 to 6. The timeline shows the temporal distribution of visited locations from Friday to Monday. Solid marks represent restaurants. Hollow marks 0, 1, 2 represent home, work place, and shopping mall, respectively. Restaurants 3, 4, 5 and 6 are functionally relevant but are temporally non-successive and spatially distanced.}
\label{fig1}
\end{figure}

With the continuously upcoming novel models pushing forward our understanding of mobility prediction, several key problems remain unsolved. 
1) First, the correlations between non-adjacent locations and non-contiguous visits have not been learned effectively. The mobility of users may depend more on relevant locations visited a few days ago rather than irrelevant locations visited just now. Moreover, it is not rare for a user to visit distanced locations that are functionally relevant/similar. In a special example shown in \textbf{Figure 1}, a user always dines at a certain restaurant near the workplace on Friday evening, go to some shopping malls on Saturday morning, and dine at a random restaurant near a mall on Saturday evening. In this case, the user has de facto made two non-consecutive visits to non-adjacent restaurants, where the explicit spatial distances between home and shopping malls and the explicit temporal interval between meals provide non-trivial information for predicting the exact location for Saturday dinner. However, most current models focused on spatial and/or temporal differences between current and future steps while ignoring spatiotemporal correlation within the trajectory.
2) Second, the previously practiced hierarchical gridding for spatial discretization is insensitive to spatial distance. The gridding-based attention network aggregates neighboring locations but cannot perceive spatial distance. Grids that are close to each other reflect no difference to those that are not, tossing a lot of spatial information. 
3) Third, previous models extensively overlooked personalized item frequency (PIF) \cite{hu2020modeling, wang2019modeling, ren2019repeatnet}. Repeated visits to the same place reflect the frequency, which emphasizes the importance of the repeated locations and the possibility of users revisiting. Previous RNN-based models and self-attention models can hardly reflect PIF due to the memory mechanism and normalization operation, respectively.

To this end, we proposed STAN, a Spatio-Temporal Self-Attention Network for the next location recommendation. In STAN\footnote{\url{https://github.com/yingtaoluo/Spatial-Temporal-Attention-Network-for-POI-Recommendation}}, we design a self-attention layer for aggregating important locations within the historical trajectory and another self-attention layer for recalling the most plausible candidates, both with the consideration of a point-to-point explicit spatiotemporal effect. Self-attention layers can assign different weights to each visit within the trajectory, which overcomes the long-term dependency problem of the commonly used recurrent layers.  The bi-layer system allows effective aggregation that considers PIF. We employ linear interpolation for the embedding of spatiotemporal transition matrix to address the sparsity problem, which is sensitive to spatial distance, unlike GPS gridding. STAN can learn correlations between non-adjacent locations and non-contiguous visits owing to the spatiotemporal effect of all check-ins fed into the model. 

To summarize, our contributions are listed as follows:
\begin{itemize}
\item We propose STAN, a spatiotemporal bi-attention model, to fully consider the spatiotemporal effect for aggregating relevant locations. To our best recollection, STAN is the first model in POI recommendation that explicitly incorporates spatiotemporal correlation to learn the regularities between non-adjacent locations and non-contiguous visits. 
\item We replace the GPS gridding with a simple linear interpolation technique for spatial discretization, which can recover spatial distances and reflect user spatial preference, instead of merely aggregating neighbors. We integrate this method into STAN for more accurate representation.
\item We specifically propose a bi-attention architecture for PIF. The first layer aggregates relevant locations within the trajectory for updated representation, so that the second layer can match the target to all check-ins, including repetition.
\item Experiments on four real-world datasets are conducted to evaluate the performances of the proposed method. The result shows that the proposed STAN outperforms the accuracy of state-of-the-art models by more than 10\%. 
\end{itemize}

\section{Related Works}
In this section, we briefly review some works on sequential recommendation and the next POI recommendation. The next POI recommendation can be viewed as a special sub-task of sequential recommendation with spatial information.

\subsection{Sequential Recommendation}
The sequential recommendation was mainly modeled by two schools of models: Markov-based models and deep learning-based models.

Markov-based models predict the probability of the next behavior via a transition matrix. Due to the sparsity of sequential data, the Markov model can hardly capture the transition of intermittent visits. Matrix factorization models \cite{koren2009matrix, rendle2010factorization} are proposed to approach this problem, with further extensions \cite{SuccessivePOI, he2016fusing} find that explicit spatial and temporal information help a lot with recommendation performance. In general, Markov-based models mainly focus on the transition probability between two consecutive visits.

Challenged by the flaws of Markov models, deep learning-based models thrive to replace them.
Among them, models based on RNN \cite{zhang2014sequential} are representative and quickly develop as strong baselines.
They have achieved satisfactory performances on variety of tasks, such as session-based recommendation \cite{hidasi2015session, li2017neural}, next basket recommendation \cite{yu2016dynamic} and next item recommendation \cite{zhou2019deep,chen2018sequential}.
Meanwhile, time intervals between adjacent behaviors are incorporated in the RNN-based recommendation models \cite{liu2016context,zhu2017next}, for better preserving the dynamic characteristics of user history.
Besides RNN, other deep learning methods are also considered.
For example, metric embedding algorithms \cite{feng2015personalized, feng2020hme}, convolutional neural networks \cite{Caser,yuan2019simple,wang2019towards}, reinforcement learning algorithms \cite{massimo2018harnessing}, and graph network \cite{wu2019session,yu2020tagnn} are proposed one by one for sequential recommendation.
Recently, researchers extensively use self-attention \cite{vaswani2017attention} for sequential recommendation, where a model named SASRec \cite{kang2018self} is proposed. Based on SASRec, time intervals within user sequence are considered \cite{li2020time,ye2020time}.
Moreover, as discussed in \cite{hu2020modeling}, Personalized Item Frequency (PIF) is very important for sequential recommendations. RNN-based sequential recommenders have been proven to be unable for effectively capturing PIF. In models based on self-attention, PIF is also hard to capture due to the normalization in attention modules. After normalization, the representation of previous histories is reduced to a single vector of embedding dimension. Matching each candidate with this representation can hardly reflect PIF information.

\subsection{Next POI Recommendation}

Most existing next POI recommendation models are based on RNN.
STRNN \cite{STRNN} uses temporal and spatial intervals between every two consecutive visits as explicit information to improve model performance, which has also been applied in public security evaluation \cite{wu2016sape}. SERM \cite{Serm} jointly learns temporal and semantic contexts that reflect user preference. DeepMove \cite{DeepMove} combines an attention layer for learning long-term periodicity with a recurrent layer for learning short-term sequential regularity and learned from highly correlated trajectories.
Regarding the use of spatiotemporal information in the next location recommendation, many previous works only used explicit spatiotemporal intervals between two successive visits in a recurrent layer. STRNN \cite{STRNN} directly uses spatiotemporal intervals between successive visits in a recurrent neural network. Then, Time-LSTM\cite{zhu2017next} proposes to add time gates to the LSTM structure to better adapt the spatiotemporal effect. STGN \cite{STGN} further enhances the LSTM structure by adding spatiotemporal gates. ATST-LSTM \cite{huang2019attention} uses an attention mechanism to assist LSTM in assigning different weights to each check-in, which starts to use attention but still only considered successive visits. LSTPM \cite{LSTPM} proposes a geo-dilated RNN that aggregates locations visited recently, but only for short-term preference. 
Inspired by sequential item recommendation \cite{kang2018self}, GeoSAN \cite{lian2020geography} uses self-attention model in next location recommendation that allows point-to-point interaction within the trajectory. However, GeoSAN ignores the explicit modeling of time intervals and spatial distances, as the gridding method for spatial discretization used in GeoSAN can not well capture the exact distances. In other words, all previous methods have not effectively considered non-trivial correlations between non-adjacent locations and non-contiguous visits.
Moreover, these models also have problems in modeling PIF information.

\section{Preliminaries}
In this section, we give problem formulations and term definitions. We denote the set of user, location and time as 
$\textit{U} = \{ u_1, u_2, ..., u_{\textit{U}} \}$, $\textit{\MakeUppercase{L}} = \{ l_1, l_2, ..., l_{\textit{L}} \}$, $\textit{\MakeUppercase{T}} = \{ t_1, t_2, ..., t_{\textit{\MakeUppercase{T}}} \}$, respectively. ~\\

\noindent\textbf{Historical Trajectory}. The trajectory of user \textit{$u_i$} is temporally ordered check-ins.
Each check-in $r_k$ within the trajectory of user \textit{$u_i$} is a tuple $(u_i, l_k, t_k)$, in which $l_k$ is the location and $t_k$ is the timestamp. Each user may have a variable-length trajectory 
$\textit{tra}( u_i ) = \{ r_1 , r_2 ,..., r_{{m}_{i}} \}$.
We transform each trajectory into a fixed-length sequence $\textit{seq}( u_i ) = \{ r_1 , r_2 ,..., r_{n} \}$, 
with $n$ as the maximum length we consider. 
If $n < m_i$, we only consider the most recent $n$ check-ins. If $n > m_i$, we pad zeros to the right until the sequence length is $n$ and mask off the padding items during calculation. ~\\

\noindent\textbf{Trajectory Spatio-Temporal Relation Matrix}.
We model time intervals and geographical distances as the explicit spatio-temporal relation between two visited locations. We denote temporal interval between $i$-th and $j$-th visits as $\Delta_{ij}^t=|t_i - t_j|$, and denote spatial distance between the GPS location of $i$-th visit and the GPS location of $j$-th visit as $\Delta_{ij}^s=Haversine(GPS_i, GPS_j)$. 
Specifically, the trajectory spatial relation matrix 
$\Delta^{s} \in \mathbb{R}^{n \times n}$ and the trajectory temporal relation matrix $\Delta^{t} \in \mathbb{R}^{n \times n}$ are separately represented as: \\
\begin{align}
\Delta^{t,s} = 
\begin{bmatrix} 
\Delta_{11}^{t,s} & \Delta_{12}^{t,s} & \dots & \Delta_{1n}^{t,s} \\[5pt]
\Delta_{21}^{t,s} & \Delta_{22}^{t,s} & \dots & \Delta_{2n}^{t,s} \\
\vdots & \vdots & \ddots & \vdots \\[2pt]
\Delta_{n1}^{t,s} & \Delta_{n2}^{t,s} & \dots & \Delta_{nn}^{t,s} \end{bmatrix}
\end{align} ~\\

\noindent\textbf{Candidate Spatio-Temporal Relation Matrix}. Besides the internal explicit relation, we also consider a next spatiotemporal matrix in the paper. It calculates the distance between each location candidate $i \in [1,L]$ and each location of the check-ins $j \in [1,n]$ as
$N_{ij}^s=Haversine(GPS_i, GPS_j)$,
and represents the time intervals between $t_{m+1}$ and $\{ t_1, t_2, ..., t_{m} \}$ that are repeated L times to expand into 2D as 
$N_{ij}^t=|t_{m+1} - t_j|$. The candidate spatial relation matrix $N^{s} \in \mathbb{R}^{L \times n}$ and the candidate temporal relation matrix $N^{t} \in \mathbb{R}^{L \times n}$ are separately represented as:\\
\begin{align}
N^{t,s} = 
\begin{bmatrix} 
N_{11}^{t,s} & N_{12}^{t,s} & \dots & N_{1n}^{t,s} \\[5pt]
N_{21}^{t,s} & N_{22}^{t,s} & \dots & N_{2n}^{t,s} \\
\vdots & \vdots & \ddots & \vdots \\[2pt]
N_{L1}^{t,s} & N_{L2}^{t,s} & \dots & N_{Ln}^{t,s} \end{bmatrix}
\end{align} ~\\

\noindent\textbf{Mobility Prediction}. 
Given the user trajectory 
$( r_1 , r_2 ,..., r_{m} )$,
the location candidates 
$L=\{ l_1, l_2, ..., l_{\textit{L}} \}$, the spatio-temporal relation matrix $\Delta^{t,s}$,
and the next spatio-temporal matrix $N^{t,s}$,
our goal is to find the desired output $l \in r_{m+1}$.

\section{The Proposed Framework}
Our proposed \textit{Spatio-Temporal Attention Network} (STAN) consists of: 1) a \textbf{multimodal embedding module} that learns the dense representations of user, location, time, and spatiotemporal effect; 2) a \textbf{self-attention aggregation layer} that aggregates important relevant locations within the user trajectory to update the representation of each check-in; 3) an \textbf{attention matching layer} that calculates softmax probability from weighted check-in representations to compute the probability of each location candidate for next location; 4) a \textbf{balanced sampler} that use a positive sample and several negative samples to compute the cross-entropy loss. The neural architecture of the proposed STAN is shown in \textbf{Figure 2}.

\begin{figure}[t]
\centering
\includegraphics[width=0.8\columnwidth]{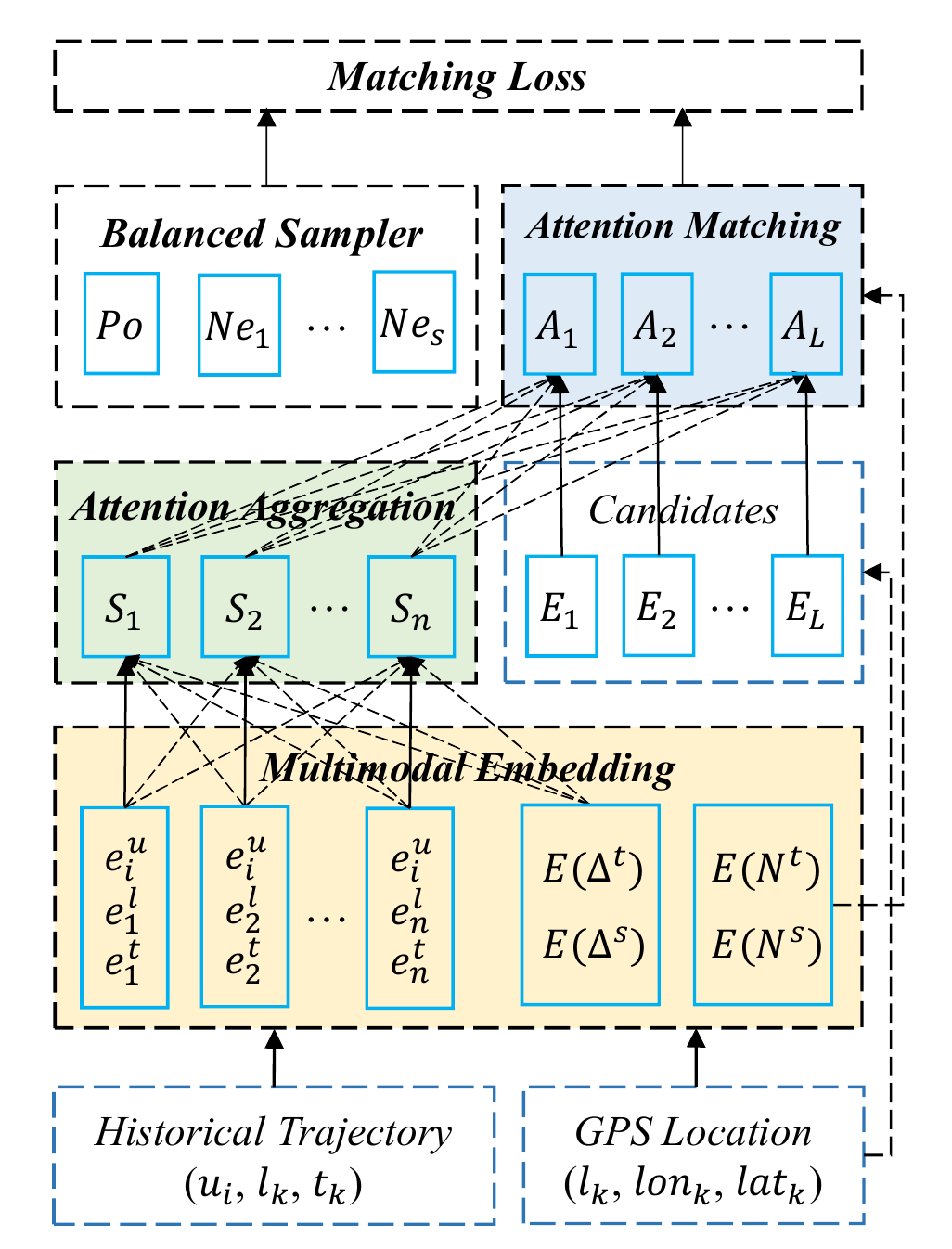}
\caption{The architecture of the proposed STAN model.}
\end{figure}

\subsection{Multimodal Embedding Module}
The multi-modal embedding module consists of two parts, namely a trajectory embedding layer and a spatio-temporal embedding layer. 
\subsubsection{User Trajectory Embedding Layer} A multi-modal embedding layer is used to encode user, location and time into latent representations. 
For user, location and time, we denote their embedded representations as 
\textit{$e^u$}$\in\mathbb{R}^{\textit{$d$}}$, 
\textit{$e^l$}$\in\mathbb{R}^{\textit{$d$}}$, 
\textit{$e^t$}$\in\mathbb{R}^{\textit{$d$}}$, respectively. 
The embedding module is incorporated into the other modules to transform the scalars into dense vectors to reduce computation and improve representation. Here, the continuous timestamp is divided by $7 \times 24 = 168$  hours that represents the exact hour in a week, which maps the original time into 168 dimensions. This temporal discretization can indicate the exact time in a day or a week, reflecting periodicity. Therefore, the input dimensions of the embeddings \textit{$e^u$}, \textit{$e^l$} and \textit{$e^t$} are U, L, and 168, respectively. The output of user trajectory embedding layer for each check-in $r$ is the sum $e^r = e^u + e^l + e^t \in \mathbb{R}^{d}$. For the embedding of each user sequence
$\textit{seq}( u_i ) = \{ r_1 , r_2 ,..., r_{n} \}$, we denote as 
$E(u_i)=\{ e^{r_1} , e^{r_2} ,..., e^{r_n} \} \in \mathbb{R}^{n \times \textit{d}}$.

\subsubsection{Spatio-Temporal Embedding Layer} A unit embedding layer is used for the dense representation of spatial and temporal differences with an hour and hundred meters as basic units, respectively. Recall that if we regard the maximum space or time intervals as the number of embeddings and discretize all the intervals, it can easily lead to a sparse relation encoding. This layer multiplies the space and time intervals each with a unit embedding vector $e_{\Delta s}$ and $e_{\Delta t}$, respectively. The unit embedding vectors reflect the continuous spatiotemporal context with the basic unit and avoid sparsity encoding with the dense dimensions. Especially, we can use this technique that is sensitive to spatial distance to replace hierarchical gridding method, which only aggregates adjacent locations and is not capable to represent spatial distance. In mathematics, the spatiotemporal difference embedding is 
$e_{ij}^{\Delta} \in \mathbb{R}^{d}$:

\begin{align}
\left\{\begin{array}{lr}
e_{ij}^{\Delta t}=\Delta_{ij}^{t} \times e_{\Delta t}
\\[5pt]
e_{ij}^{\Delta s}=\Delta_{ij}^{s} \times e_{\Delta s}
\end{array}\right.
\end{align}\\

Inspired by \cite{STRNN,liu2017multi,liu2020empirical}, we may also consider an alternative interpolation embedding layer that sets a upper-bound unit embedding vector and a lower-bound unit embedding vector and represents the explicit intervals as a linear interpolation, which is an approximation to the unit embedding layer. In experiments, the two methods have similar efficiency. The interpolation embedding is calculated as:

\begin{align}
\left\{\begin{array}{lr}
e_{ij}^{\Delta t}=
\dfrac{e_{\Delta t}^{sup}(Upper(\Delta t)-\Delta t) + e_{\Delta t}^{inf}(\Delta t-Lower(\Delta t))}
{Upper(\Delta t)-Lower(\Delta t)}
\\[12pt]
e_{ij}^{\Delta s}=
\dfrac{e_{\Delta s}^{sup}(Upper(\Delta s)-\Delta s) + e_{\Delta s}^{inf}(\Delta s-Lower(\Delta s))}
{Upper(\Delta s)-Lower(\Delta s)}
\\[6pt]
\end{array}\right.
\end{align}\\

This layer processes two matrices: the trajectory spatio-temporal relation matrix and the candidate spatio-temporal relation matrix, as described in preliminaries. Their embeddings are
$E(\Delta^t) \in \mathbb{R}^{n \times n \times d}$,
$E(\Delta^s) \in \mathbb{R}^{n \times n \times d}$,
$E(N^t) \in \mathbb{R}^{L \times n \times d}$,
and $E(N^s) \in \mathbb{R}^{L \times n \times d}$.
We can use a weighted sum of the last dimension and add spatial and temporal embeddings together to create:

\begin{align}
\left\{\begin{array}{lr}
E(\Delta)=Sum(E(\Delta^t))+Sum(E(\Delta^s)) \in \mathbb{R}^{n \times n}\\[6pt]
E(N)=Sum(E(N^t))+Sum(E(N^s)) \in \mathbb{R}^{L \times n}
\end{array}\right.
\end{align}\\

\subsection{Self-Attention Aggregation Layer} 
Inspired by self-attention mechanisms, we propose an extensional module to consider the different spatial distances and time intervals between two visits in a trajectory. This module aims at aggregating relevant visited locations and updating the representation of each visit. Self-attention layer can capture long-term dependency and assign different weights to each visit within the trajectory. This point-to-point interaction within the trajectory allows the layer to assign more weights to relevant visits. Moreover, we can easily incorporate the explicit spatio-temporal intervals into the interaction. Given the user embedded trajectory matrix $E(u)$ with non-padding length $m'$ and the spatio-temporal relation matrices $E(\Delta)$, this layer firstly construct a mask matrix $M \in \mathbb{R}^{n \times n}$ with upper left elements $\mathbb{R}^{m' \times m'}$ being ones and other elements being zeros. Then the layer computes a new sequence $S$ after converting them through distinct parameter matrices $W_Q, W_K, W_V \in \mathbb{R}^{\textit{d} \times \textit{d}}$ as

\begin{align}S(u)=Attention(E(u)W_Q, E(u)W_K, E(u)W_V, E(\Delta), M)\end{align}
with 
\begin{align}Attention(Q,K,V,\Delta, M)=
\left(M*softmax(\dfrac{QK^T+\Delta}{\sqrt{d}})\right)V\end{align}\\

Here, only the mask and softmax attention are multiplied element by element, while others use matrix multiplication. It is very important for us to consider causality that only the first $m'$ visits in the trajectory are fed into the model while predicting the $(m'+1)$-st location. Therefore, during training, we use all the $m' \in [1,m]$ to mask the input sequence and accordingly to the selected label. We can get 
$S(u) \in \mathbb{R}^{n \times \textit{d}}$ as the updated representation of the user trajectory. Another alternative implementation is to feed explicit spatio-temporal intervals into both $E(u)W_K$ and $E(u)W_V$, as TiSASRec \cite{li2020time} did. However, in experiments, we found out the two methods have similar performances. Our implementation is in a more concise form using only matrix multiplication instead of element-wise calculation.

\subsection{Attention Matching Layer} 
This module aims at recalling the most plausible candidates from all the L locations by matching with the updated representation of the user trajectory. Given the updated trajectory representation 
$S(u) \in \mathbb{R}^{n \times \textit{d}}$,
the embedded location candidates $E(l)=\{e^l_1, e^l_2, ..., e^l_L\} \in \mathbb{R}^{L \times \textit{d}}$,
and the embedding of the candidate spatio-temporal relation matrix $E(N) \in \mathbb{R}^{L \times n}$,
this layer computes the probability of each location candidate to be the next location as

\begin{align}A(u)=Matching(E(l),S(u),E(N))\end{align}
with
\begin{align}Matching(Q,K,N)=Sum\left(softmax\left(\dfrac{QK^T+N}{\sqrt{d}}\right)\right)\end{align}\\

Here, the $Sum$ operation is a weighted sum of the last dimension, converting the dimension of $A(u)$ to be $\mathbb{R}^L$. In Eq.(8), we show that the updated representations of check-ins all participate in the matching of each candidate location, unlike other self-attention models that reduce the PIF information. This is due to the design of a bi-layer system that firstly aggregates relevant locations and then recalls from representations with consideration of PIF.

\subsection{Balanced Sampler} 
Due to the unbalanced scale of positive and negative samples in $A(u)$, optimizing the cross-entropy loss is no longer efficient as the loss weights little on the momentum to push forward the correct prediction. It would be normal to observe that as the loss goes down, the recall rate also goes down. Given the user $i$'s sequence $seq(u_i)$, the matching probability of each candidate location $a_j \in A(u_i)$ for $j \in [1,L]$, and the label $l_{k}$ with number of order $k$ in the location set $L$, the ordinary cross-entropy loss is written as:
\begin{align}
-\sum_{i}\sum_{m_i}\left(log \sigma(a_k) + \sum_{j=1, j \neq k}^L log (1-\sigma(a_j))\right)
\end{align}
In this form, for every positive sample $a_k$, we need to compute $L-1$ negative samples in the meantime. Other implementations also extensively used binary cross-entropy loss that computes only one negative sample along with a positive sample. However, this may also leave many non-label samples unused throughout the entire training. Here, we can simply set the number of negative samples used in cross-entropy loss as a hyperparameter s. Here we propose a balanced sampler for randomly sampling negative samples at each step of training. Consequently, we update the random seed for the negative sampler after each training step. The loss is calculated as
\begin{align}
-\sum_{i}\sum_{m_i}\left(log \sigma(a_k) + 
\sum_{\substack{(j_1, j_2, ..., j_s) \in [1,L] \\ (j_1, j_2, ..., j_s) \neq k}} log (1-\sigma(a_j))\right)
\end{align}

\section{Experiments}
In this section, we show our empirical results to make a fair comparison with other models quantitatively. We show a table of datasets, a table of recommendation performance under the evaluation of top$k$ recall rates, figures of model stability, and the visualization of attention weights in STAN aggregation.

\subsection{Datasets}
We evaluate our proposed STAN model on four real-world datasets: Gowalla \footnote{\url{http://snap.stanford.edu/data/loc-gowalla.html}}, SIN \footnote{\url{https://www.ntu.edu.sg/home/gaocong/data/poidata.zip}}, TKY and NYC \footnote{\url{http://www-public.imtbs-tsp.eu/~zhang_da/pub/dataset_tsmc2014.zip}}. The numbers of users, locations, and check-ins in each dataset are shown in \textbf{Table 1}. In experiments, we use the original raw datasets that only contain the GPS of each location and user check-in records, and pre-process them following each work's protocol. In regard to the pre-processing technique of datasets, many previous works used sliced trajectory with a fixed-length window or maximum time interval. We follow each work's setup, although this could prevent the model from learning long-time dependency. For each user that has $m$ check-ins, we divide a dataset into training, validation, and test datasets. The number of training set is $m-3$, with the first $m' \in [1,m-3]$ check-ins as input sequence and the $[2,m-2]$-nd visited location as label; the validation set uses the first $m-2$ check-ins as input sequence and the $(m-1)$-st visited location as label; the test set uses the first $m-1$ check-ins as input sequence and the $m$-th visited location as label. The split of datasets follows the causality that no future data is used in the prediction of future data.

\begin{table}
  \caption{Basic dataset statistics.}
  \label{tab:freq}
  \begin{tabular}{ccccl}
    \toprule
    \  & Gowalla & TKY & SIN & NYC\\
    \midrule
    \#users & 53008 & 2245 & 2032 & 1064\\
    \#locations & 121944 & 7872 & 3662 & 5136\\
    \#check-ins & 3302414 & 447571 & 179721 & 147939\\
  \bottomrule
\end{tabular}
\end{table}

\begin{table*}[]
\caption{Recommendation performance comparison with baselines.}
\begin{tabular}{cccccccccccc}
\hline
            & \multicolumn{2}{c}{Gowalla}       &           & \multicolumn{2}{c}{TKY}           &           & \multicolumn{2}{c}{SIN}           &           & \multicolumn{2}{c}{NYC}           \\ \cline{2-3} \cline{5-6} \cline{8-9} \cline{11-12} 
            & Recall@5        & Recall@10       &           & Recall@5        & Recall@10       &           & Recall@5        & Recall@10       &           & Recall@5        & Recall@10       \\ \hline
STRNN       & 0.1664          & 0.2567          &           & 0.1836          & 0.2791          &           & 0.1791          & 0.2016          &           & 0.2365          & 0.2802          \\
DeepMove    & 0.1959          & 0.2699          &           & 0.2684          & 0.3509          &           & 0.2389          & 0.3155          &           & 0.3268          & 0.4014          \\
STGN        & 0.1528          & 0.2422          &           & 0.1940          & 0.2710          &           & 0.2292          & 0.2727          &           & 0.2439          & 0.3015          \\
ARNN        & 0.1810          & 0.2745          &           & 0.1852          & 0.2696          &           & 0.1817          & 0.2538          &           & 0.1970          & 0.3483          \\
LSTPM       & 0.2015          & 0.2701          &           & 0.2568          & 0.3310          &           & 0.2579          & 0.3327          &           & 0.2791          & 0.3564          \\
TiSASRec    & 0.2411          & 0.3546          &           & 0.3031          & 0.3693          &           & 0.2963          & 0.3753          &           & 0.3664          & 0.5020          \\
GeoSAN      & 0.2764          & 0.3645          &           & 0.2957          & 0.3740          &           & 0.3397          & 0.3943          &           & 0.4006          & 0.5267          \\
STAN        & \textbf{0.3016} & \textbf{0.3998} & \textbf{} & \textbf{0.3461} & \textbf{0.4264} & \textbf{} & \textbf{0.3751} & \textbf{0.4301} & \textbf{} & \textbf{0.4669} & \textbf{0.5962} \\ \hline
Improvement & 9.12\%          & 9.68\%          &           & 17.04\%         & 14.01\%         &           & 10.42\%         & 9.08\%          &           & 16.55\%         & 13.20\%         \\ \hline
\end{tabular}
\end{table*}

\subsection{Baseline Models}
We compare our STAN with the following baselines:
\begin{itemize}
\item \textbf{STRNN} \cite{STRNN}: an invariant RNN model that incorporates spatio-temporal features between consecutive visits.
\item \textbf{DeepMove} \cite{DeepMove}: a state-of-the-art model with recurrent and attention layers to capture periodicity.
\item \textbf{STGN}  \cite{STGN}: a state-of-the-art model that adds time and distance interval gates to LSTM.
\item \textbf{ARNN} \cite{ARNN}: a state-of-the-art model that uses semantic and spatial information to construct knowledge graph and improve the performance of sequential LSTM model.
\item \textbf{LSTPM} \cite{LSTPM}: a state-of-the-art model that combines long-term and short-term sequential models for recommendation.
\item \textbf{TiSASRec} \cite{li2020time}: a state-of-the-art model that uses self-attention layers with explicit time intervals for sequential recommendation, but it uses no spatial information.
\item \textbf{GeoSAN} \cite{lian2020geography}: a state-of-the-art model that uses hierarchical gridding of GPS locations for spatial discretization and uses self-attention layers for matching, without use of explicit spatio-temporal interval.
\end{itemize}

\subsection{Evaluation Matrices}
We adopt the top$k$ recall rates, Recall@5 and Recall@10, to evaluate recommendation performance. Recall@k counts the rate of true positive samples in all positive samples, which in our case means the rate of the label in the top$k$ probability samples. For evaluation, we drop the balanced sampler module and directly recall the target from A, the output of the attention matching layer. The larger the Recall@k, the better the performance.

\subsection{Settings}
There are two kinds of hyperparameters: (i) common hyperparameters that are shared by all models; (ii) unique hyperparameters that depend on each model's framework. We train the common hyperparameters on a simple recurrent neural network and then apply them to all models, which helps reduce the training burden. The embedding dimension $d$ to 50 for TKY, SIN and NYC datasets and 10 for gowalla dataset. We use the Adam optimizer with default betas, the learning rate of 0.003, the dropout rate of 0.2, the training epoch of 50, and the maximum length for trajectory sequence of 100. Fixing these common hyperparameters, we fine-tune the unique hyperparameters for each model. In our model, the number of negative samples in the balanced sampler is optimal at 10. 

\subsection{Recommendation Performance}
\textbf{Table 2} shows the recommendation performance of our model and baselines on the four datasets. All the differences between different methods are statistically significant ($error < 5e^{-5}$). We use a T-test with a p-value of 0.01 to evaluate the performance improvement provided by STAN. Here, we use the averaged performance run by 10 times and reject the H0 hypothesis. Therefore, we know the improvement of STAN is statistically significant. 

We can see that our model unequivocally outperforms all compared models with 9\%-17\% improvement in recall rates. We show in \textbf{Figures 3 and 4} that the model is stable under hyperparameter tuning. Among baseline models, self-attention models such as TiSASRec and GeoSAN clearly have better performances over RNN-based models. It is not a surprise since previous RNN-based models often use sliced short trajectories instead of long trajectories, which tossed long-term periodicity and can hardly capture the exact influence of each visits towards the next movement. It should be noted that we do not use any semantic information to construct knowledge graph to perform meta-path in ARNN, as semantic analysis was not performed by other baselines in the comparison.

Among RNN-based models, LSTPM and DeepMove have relatively better performances, due to their consideration of periodicity. Among self-attention models, TiSASRec used temporal intervals and GeoSAN considered geographical partitions. Only STAN fully considers the spatio-temporal intervals within the sequences for modeling non-consecutive visits and non-adjacent locations, and modifies attention architecture to adapt PIF information instead of inheriting the transformer \cite{vaswani2017attention} structure directly. In addition, because STRNN and TiSASRec both use temporal intervals, we can compare their performances to evaluate the improvement provided by self-attention modules versus recurrent layers.

We can also refer to \textbf{Table 3}, where the $-ALL$ model represents a variant STAN model without spatio-temporal intervals and the balanced sampler. $-ALL$ model is different from ordinary self-attention models 
only on the bi-layer system, which considers PIF information. $-ALL$ has a slightly worse performance than GeoSAN on the recall rates of the four datasets, but is slightly better than TiSASRec and much better than RNN-based models. This tells us that the bi-layer system which considers PIF is approximately as important as time intervals incorporated into the attention systems.

\begin{table*}[]
\caption{Ablation Analysis, in which we compare different modules in STAN.}
\begin{tabular}{cccccccccccc}
\hline
       & \multicolumn{2}{c}{Gowalla}       &           & \multicolumn{2}{c}{TKY}           &           & \multicolumn{2}{c}{SIN}           &           & \multicolumn{2}{c}{NYC}           \\ \cline{2-3} \cline{5-6} \cline{8-9} \cline{11-12} 
       & Recall@5        & Recall@10       &           & Recall@5        & Recall@10       &           & Recall@5        & Recall@10       &           & Recall@5        & Recall@10       \\ \hline
STAN   & \textbf{0.3016} & \textbf{0.3998} & \textbf{} & \textbf{0.3461} & \textbf{0.4264} & \textbf{} & \textbf{0.3751} & \textbf{0.4301} & \textbf{} & \textbf{0.4669} & \textbf{0.5962}  \\
-TIM-BS  & 0.2835          & 0.3718          &           & 0.3006          & 0.3819          &           & 0.3416          & 0.3873          &           & 0.4126          & 0.5245          \\
-EWTI-BS    & 0.2794          & 0.3717          &           & 0.3052          & 0.3781          &           & 0.3404          & 0.3890          &           & 0.4083          & 0.5272      \\
-TIM & 0.2946          & 0.3925          &           & 0.3315          & 0.4099          &           & 0.3643          & 0.4176          &           & 0.4495          & 0.5814          \\
-SIM-BS    & 0.2823          & 0.3729          &           & 0.3123          & 0.3865          &           & 0.3337          & 0.3901          &           & 0.4126          & 0.5299          \\
-EWSI-BS  & 0.2812          & 0.3724          &           & 0.3132          & 0.3794          &           & 0.3313          & 0.3916          &           & 0.4124          & 0.5277          \\
-SIM & 0.2977          & 0.3908          &           & 0.3405          & 0.4141          &           & 0.3636          & 0.4165          &           & 0.4502          & 0.5860          \\
-ALL   & 0.2645          & 0.3531          &           & 0.2867          & 0.3660          &           & 0.3239          & 0.3776          &           & 0.3896          & 0.5094          \\ \hline
\end{tabular}
\end{table*}

\begin{figure}[t]
\caption{Impact of embedding dimension.}
\begin{minipage}[t]{0.45\linewidth}
\centering
\includegraphics[width=1\linewidth]{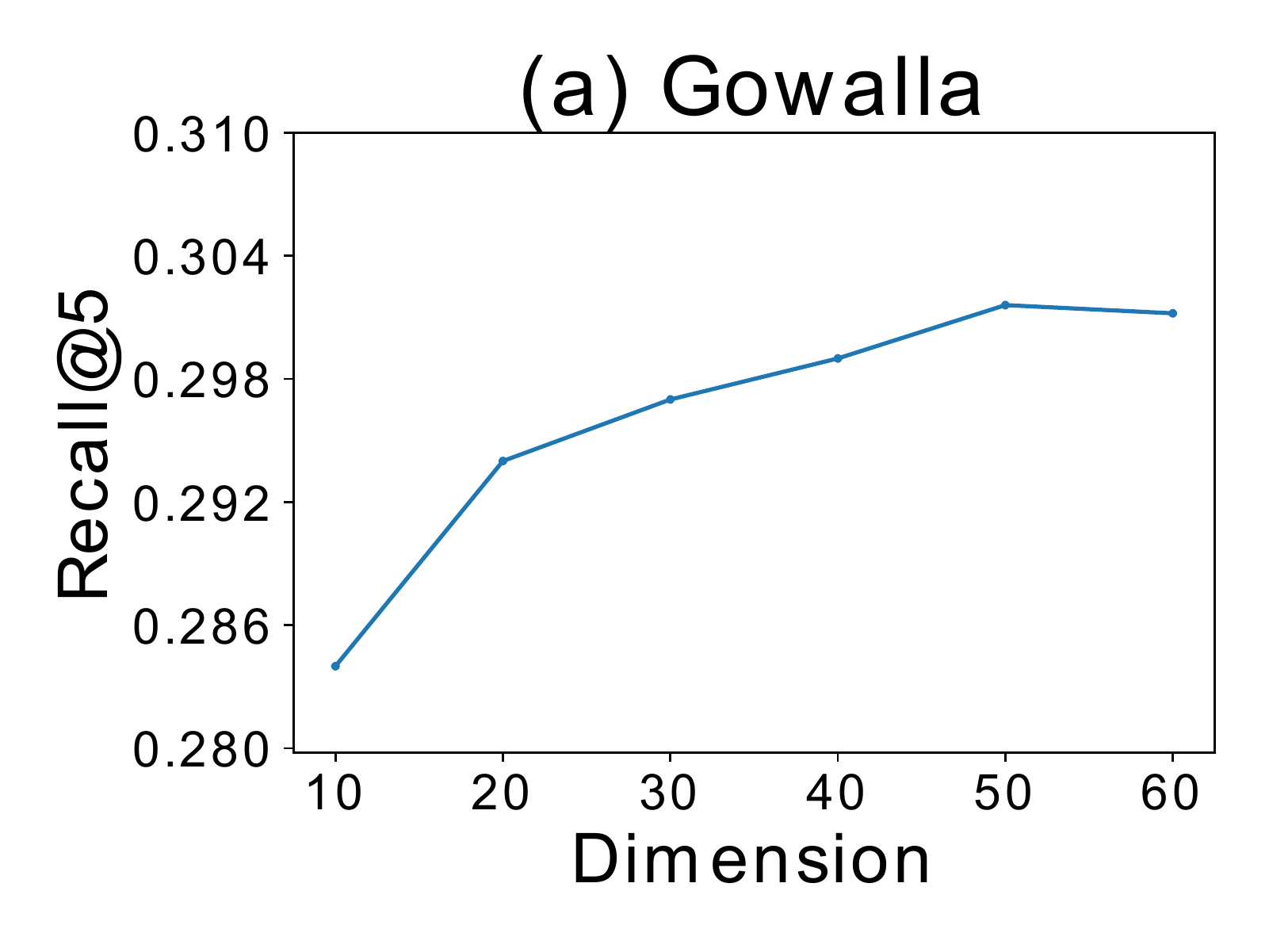}
\end{minipage}%
\begin{minipage}[t]{0.45\linewidth}
\centering
\includegraphics[width=1\linewidth]{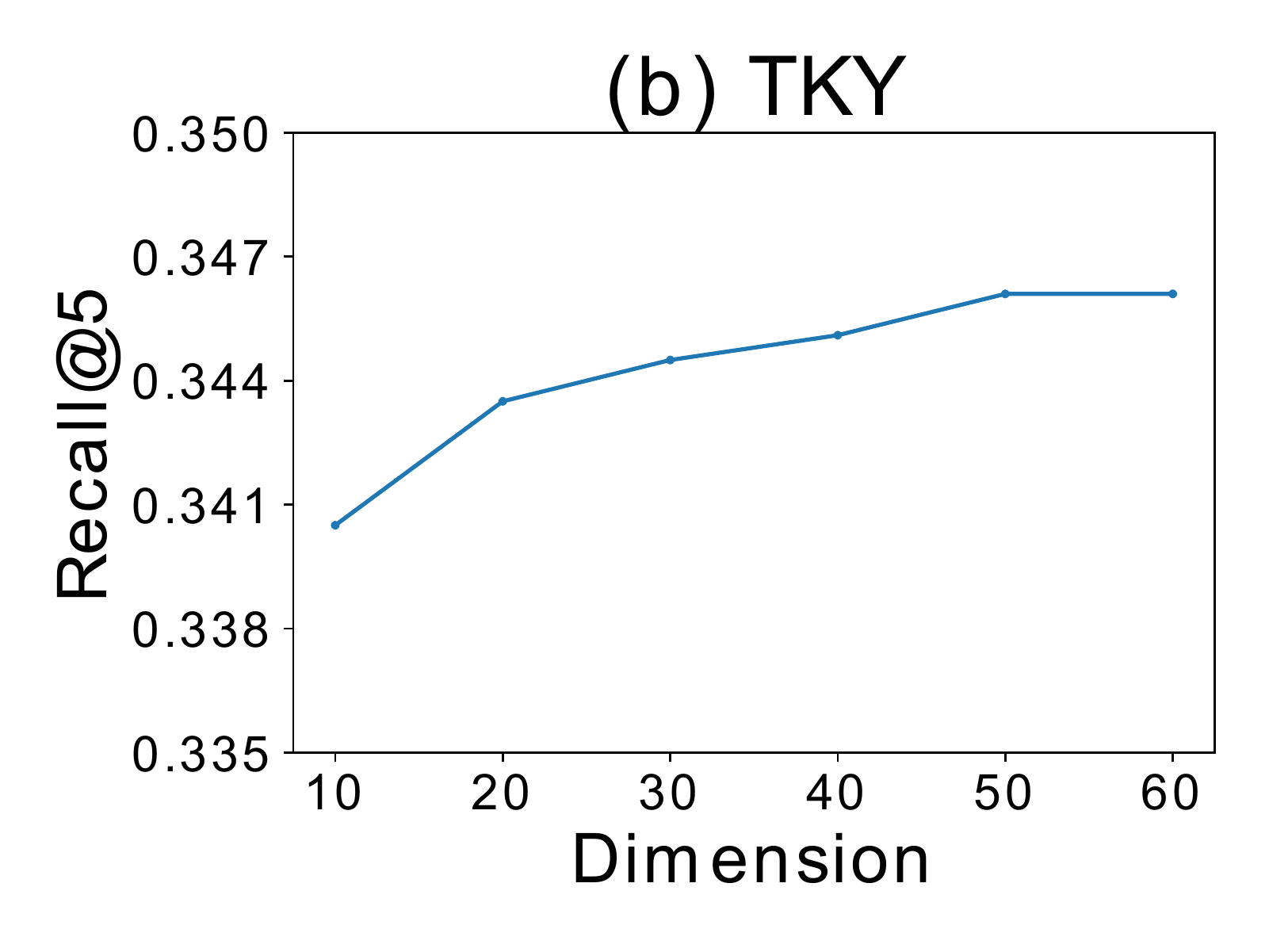}
\end{minipage}\\
\begin{minipage}[t]{0.45\linewidth}
\centering
\includegraphics[width=1\linewidth]{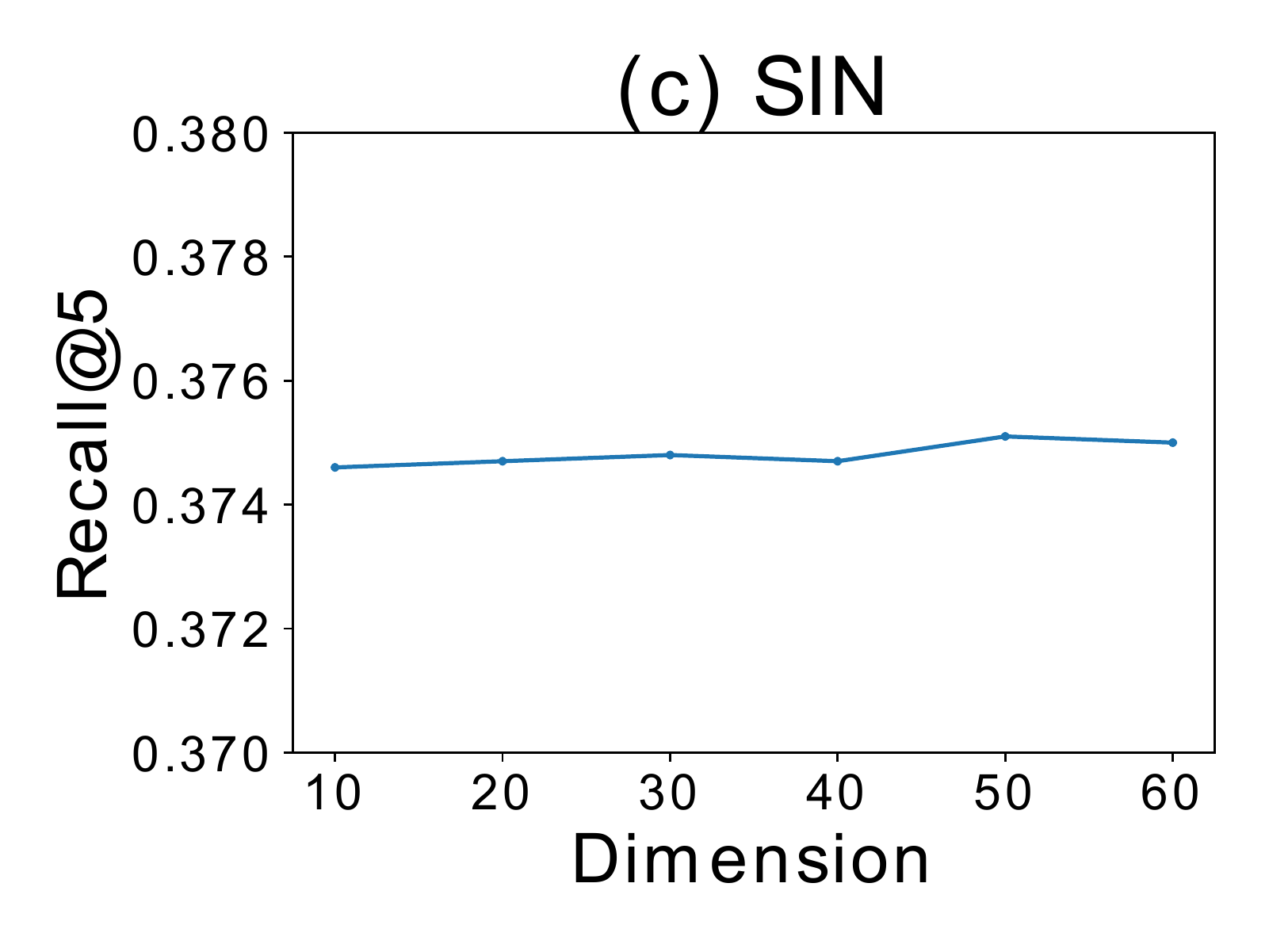}
\end{minipage}
\begin{minipage}[t]{0.45\linewidth}
\centering
\includegraphics[width=1\linewidth]{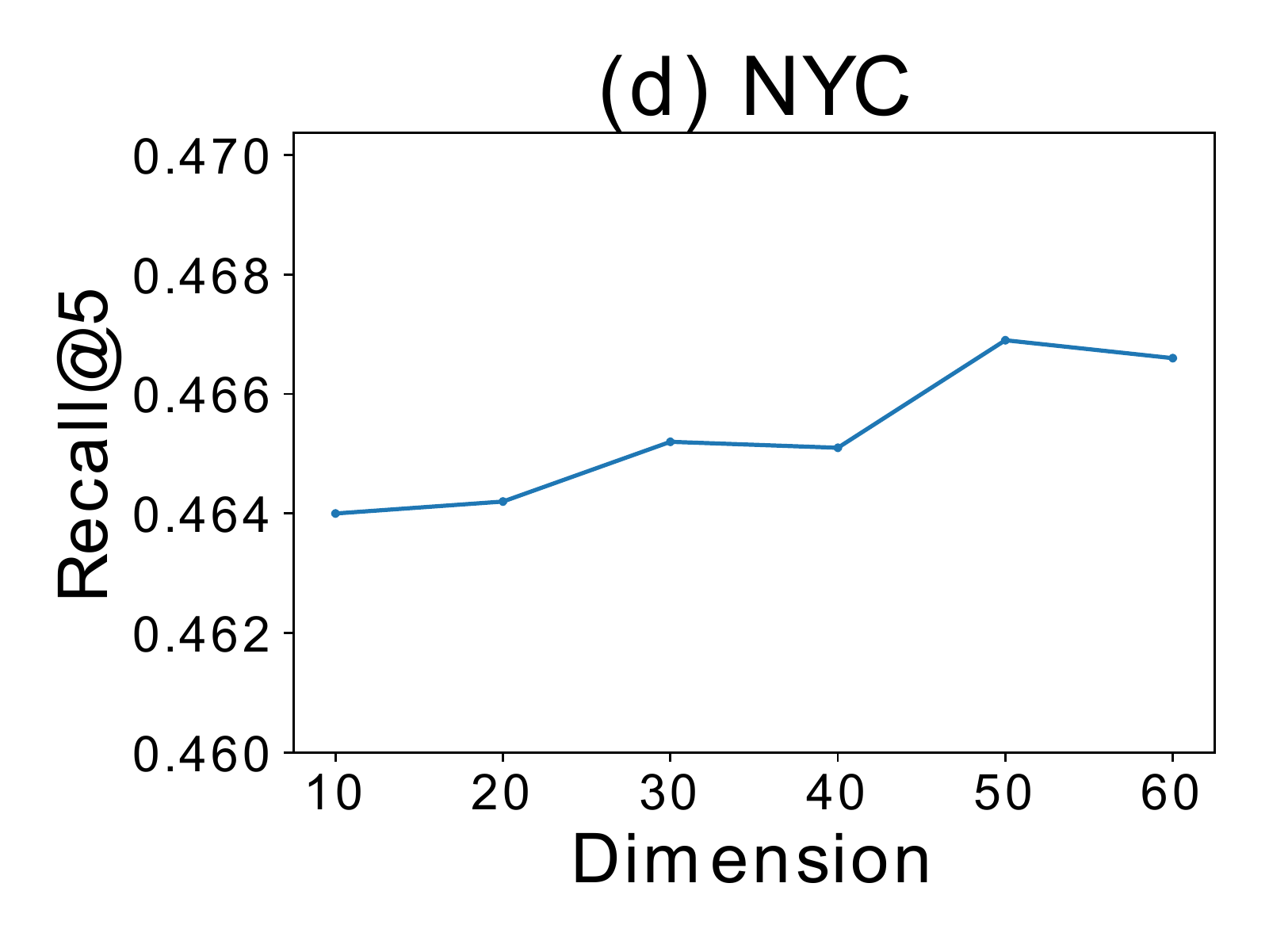}
\end{minipage}
\end{figure}

\subsection{Ablation Study}
To analyze different modules in our model, we conduct an ablation study in this section. We denote the based model as STAN, with spatio-temporal intervals and a balanced sampler. We drop different components to form variants. The components are listed as:
\begin{itemize}
\item SIM (Spatial Intervals in Matrix): This denotes the explicit spatial intervals we use within the trajectory as a matrix.
\item EWSI (Element-Wise Spatial Intervals): This denotes the element-wise spatial intervals following the structure of TiSASRec \cite{li2020time}.
\item TIM (Temporal Intervals in Matrix): This denotes the explicit temporal intervals we use within the trajectory as a matrix.
\item EWTI (Element-Wise Temporal Intervals): This denotes the element-wise temporal intervals following the structure of TiSASRec \cite{li2020time}.
\item BS (Balanced Sampler): Balanced sampler for calculating loss. 
\end{itemize}

\textbf{Table 3} shows the results of the ablation study. We find that a balanced sampler is crucial for improving the recommendation performance, which provides a nearly 5-12\% increase in recall rates. Spatial and temporal intervals can explicitly express the correlation between non-consecutive visits and non-adjacent locations. Adding spatial distances and temporal intervals all provide nearly 4-8\% increase in recall rates. We also find that our method to introduce spatio-temporal correlations is equivalent to the method used in TiSASRec \cite{li2020time}, while our method is easier to implement and can be computationally convenient due to its matrix form. The worst condition is that none of the spatio-temporal intervals nor balanced sampler is used, in which the Recall@5 and Recall@10 decrease drastically. Even so, this $-ALL$ ablated model still outperforms previously reported RNN-based models such as DeepMove, STRNN, and STGN. $-ALL$ model with the bi-layer system can consider PIF information. This explains why $-ALL$ still has a better performance over TiSASRec and RNN-based models. This tells us that the bi-layer system which considers PIF is as important as time intervals incorporated into self-attention systems.

\begin{figure}[t]
\caption{Impact of number of negative samples.}
\begin{minipage}[t]{0.45\linewidth}
\centering
\includegraphics[width=1\linewidth]{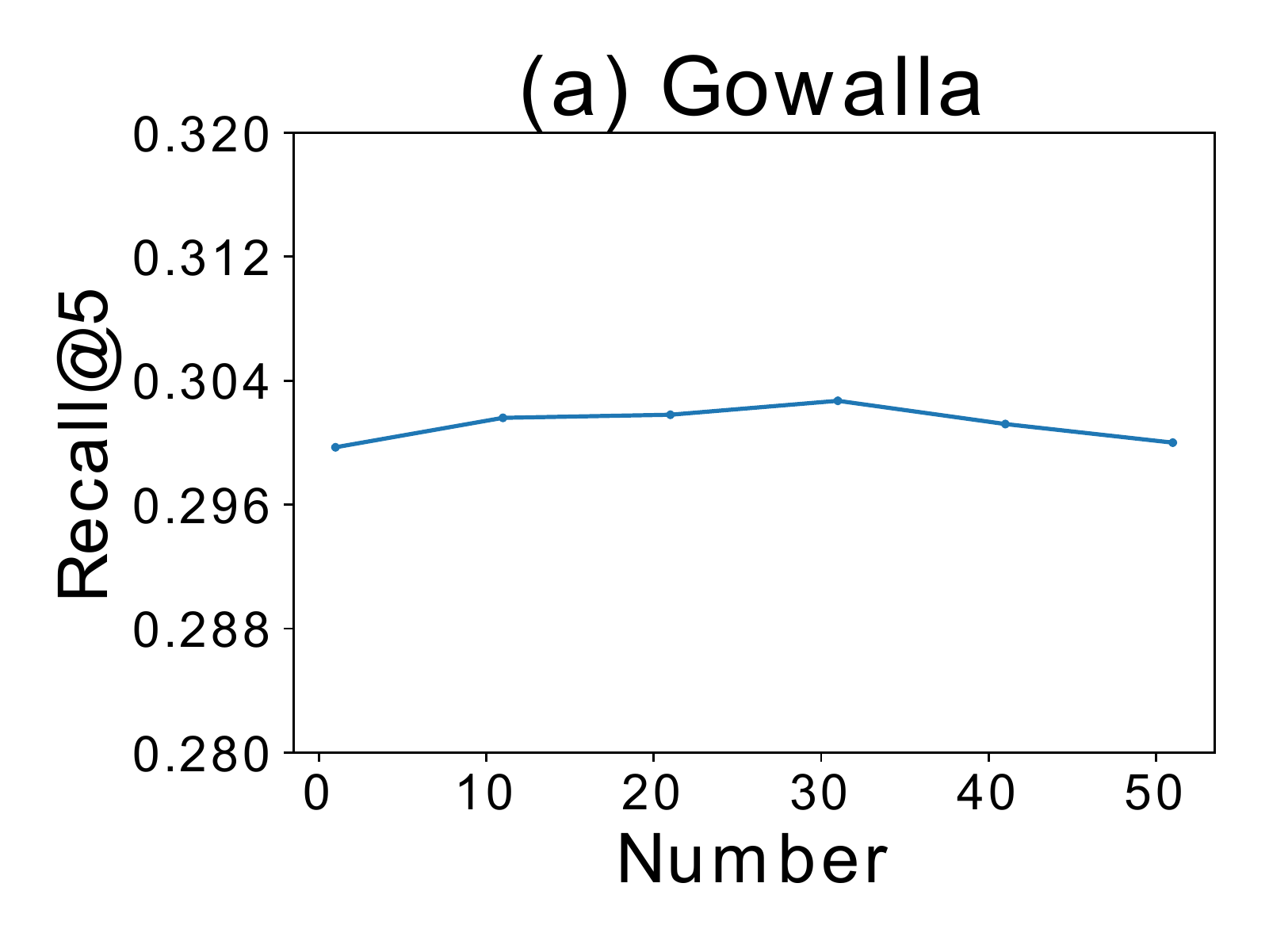}
\end{minipage}%
\begin{minipage}[t]{0.45\linewidth}
\centering
\includegraphics[width=1\linewidth]{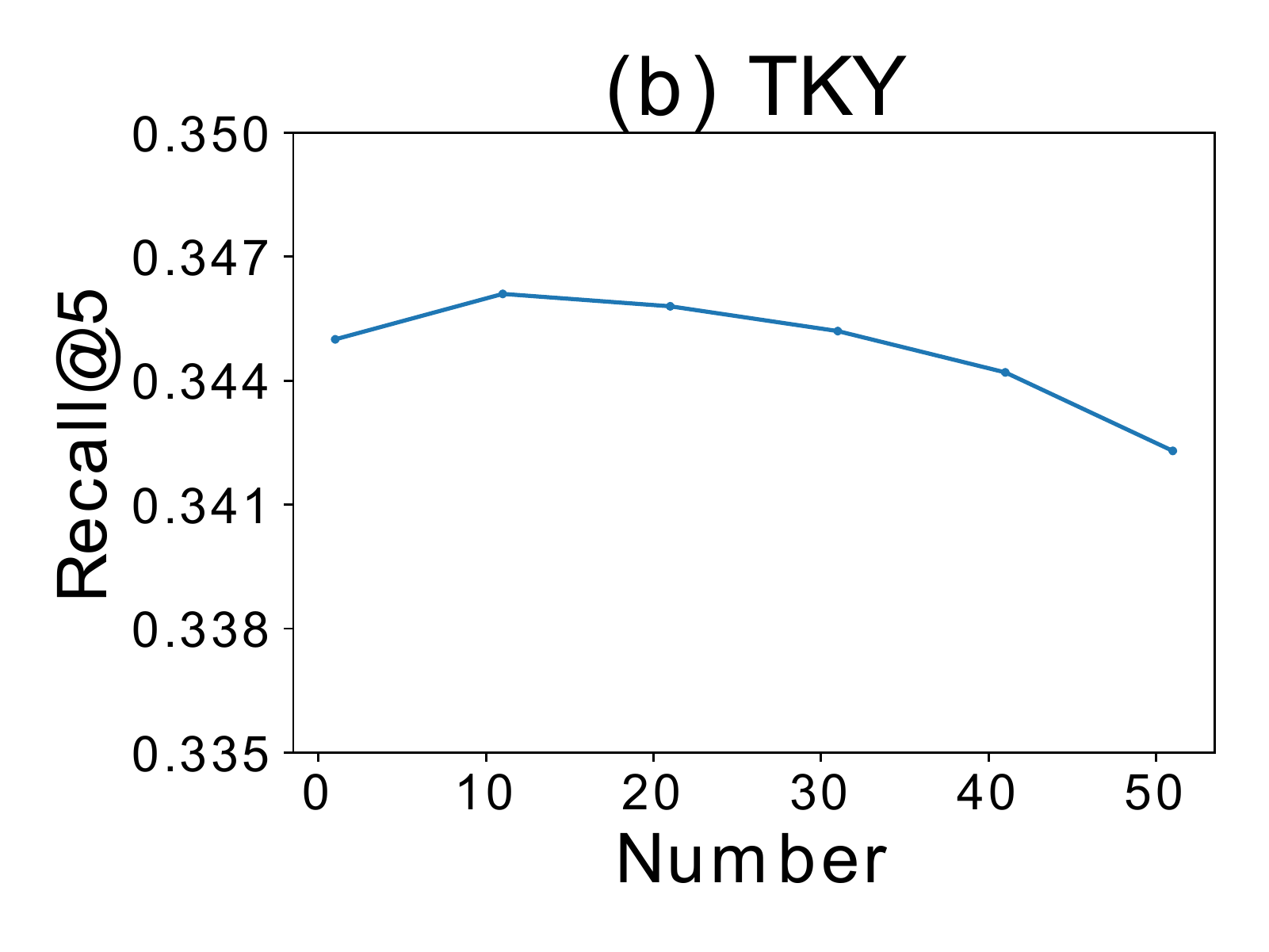}
\end{minipage}\\
\begin{minipage}[t]{0.45\linewidth}
\centering
\includegraphics[width=1\linewidth]{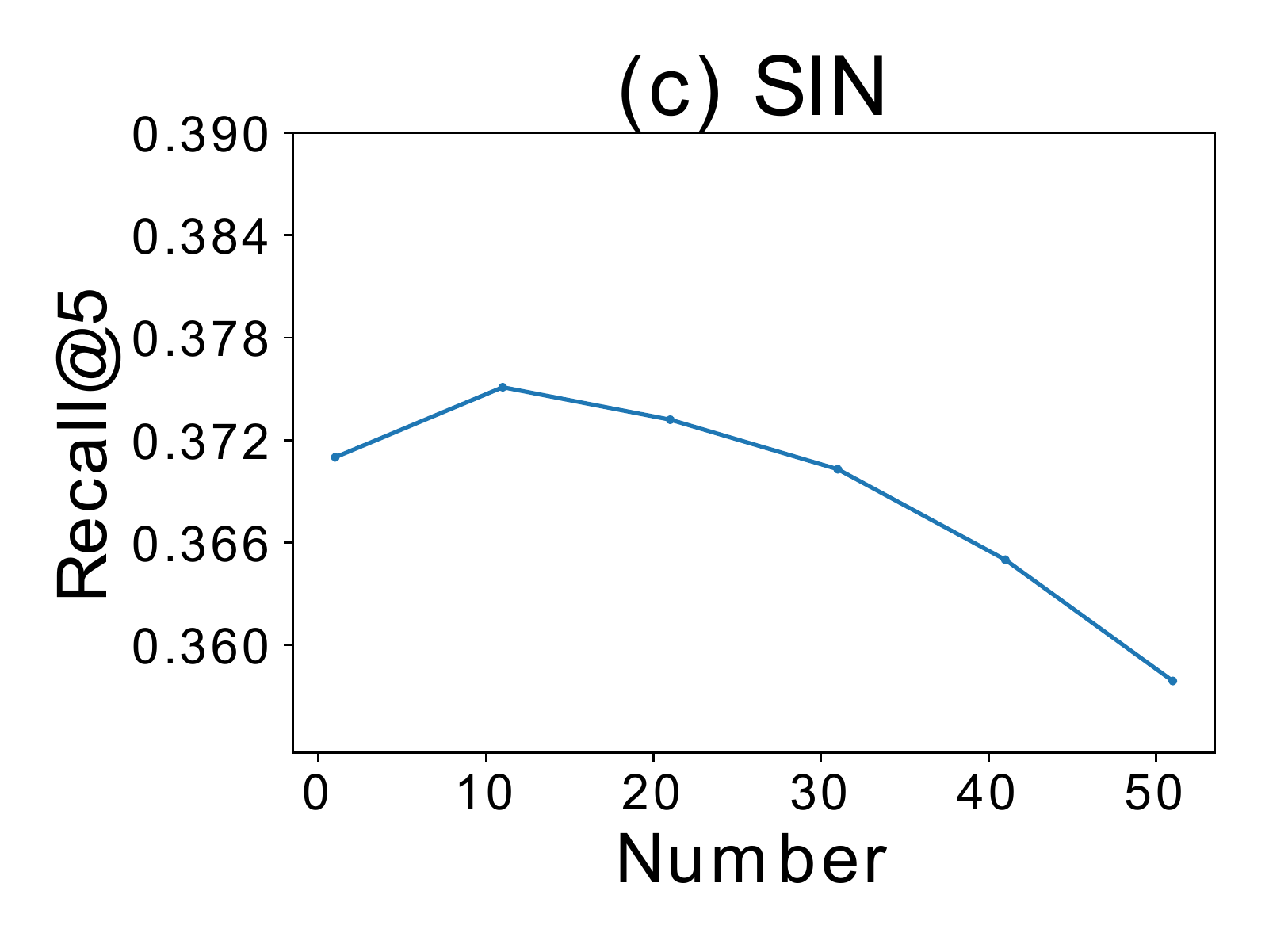}
\end{minipage}
\begin{minipage}[t]{0.45\linewidth}
\centering
\includegraphics[width=1\linewidth]{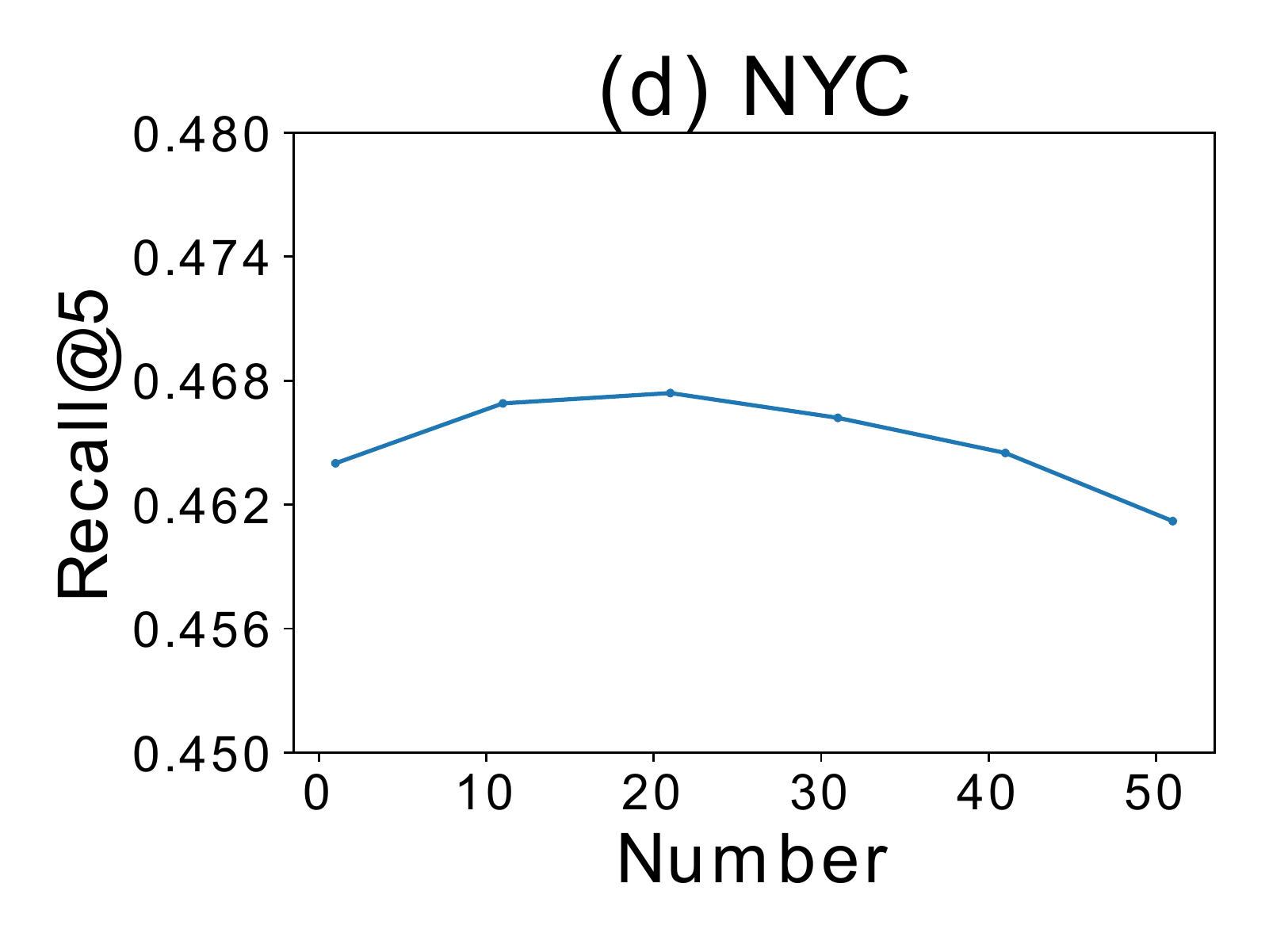}
\end{minipage}
\end{figure}

\subsection{Stability Study}
\subsubsection{Embedding dimension} We vary the dimension of embedding $d$ in the multimodal embedding module from 10 to 60 with step 10. \textbf{Figure 3} shows that $d=50$ is the best dimension for trajectory and spatio-temporal embedding. In general, the recommendation performance of our model is insensitive to the hyperparameter $d$, with less than 6\% change rate for the Gowalla dataset and less than 2\% change rate for other datasets. As long as $d$ is large than 30, the change in recommendation performance will be less than 0.5\%, which can be ignored.

\subsubsection{Number of negative samples} We experiment a series of number of negative samples $s=[1, 10, 20, 30, 40, 50]$ in the balanced sampler. \textbf{Figure 4} shows that the number of negative samples less than 20 can all produce stable recommendations for all datasets. STAN is specifically insensitive to the number of negative samples for the Gowalla dataset, which has as many as 121944 locations. This indicates that the larger the dataset, the larger the optimal number of negative samples. As the number of negative samples increases, the balanced loss will tend to the ordinary cross-entropy loss. In \textbf{Table 3}, we found that the balanced sampler is crucial for improving recommendation performance. If the number of negative samples is above the threshold, the recall rate will drop drastically.

\subsection{Interpretability Study}
To understand the mechanism of STAN, the aggregation of non-consecutive visits and non-adjacent locations performed by the self-attention aggregation layer is at the core. We visualize the correlation matrix $Cor$ of the attention weights in \textbf{Figure 5}. Each element $Cor_{i,j}$ of the matrix represents the weighted influence of $j$-th visited location on $i$-th visited location. The correlation matrix is calculated as the softmax of the multiplication of query and key in the self-attention aggregation layer. The value of each element in this correlation matrix is either tending to 1 or 0, as a result of softmax operation. Using the correlation matrix to times the original check-in embeddings, we can update the representations of the trajectory. \textbf{Figure 5} is based on a slice of real user trajectory example that is discussed in \textit{Introduction Section} and \textbf{Figure 1}. 

Here, different locations are classified and named by numbers from 0 to 6. By query of the exact GPS, we find that locations 0, 1, 2 are home, workplace, and shopping mall, respectively. Locations 3, 4, 5 and 6 are restaurants. Figure 5(a) shows the spatial correlation of visited locations that is attained by Figure 5(b), where locations with the yellow-colored marks and locations within the range of the same dark circles are aggregated together. This shows that not only adjacent locations but also non-adjacent locations are correlated. Locations 3, 4, 5 and 6 are all restaurants and are often visited at the exact time for meals. We can tell from the correlation matrix that they are relevant, despite that they are spatially distanced. The temporal order of this trajectory example is shown in the timeline of \textbf{Figure 1}. This is a sliced sparse trajectory as we edit off the irrelevant visits to focus on the correlation of restaurants. The time and order of these restaurants being visited are not consecutive but are still aggregated together. This gives evidence that visited temporally non-consecutive locations may be correlated. Both shreds of evidence in space and time demonstrate our motivation.

\section{Conclusion}
In this work, we propose a spatio-temporal attention network, abbreviated as STAN. We use a real trajectory example to illustrate the functional relevance between non-adjacent locations and non-consecutive visits, and propose to learn the explicit spatio-temporal correlations within the trajectory using a bi-attention system. This architecture firstly aggregates spatio-temporal intervals within the trajectory and then recalls the target. Because all the representations of the trajectory are weighted, the recall of the target fully considers the effect of personalized item frequency (PIF). We propose a balanced sampler for matching calculating cross-entropy loss, which outperforms the commonly practiced binary and/or ordinary cross-entropy loss. We perform comprehensive ablation study, stability study, and interpretability study in the experimental section. We prove an improvement of recall rates by the proposed components and very robust stability against hyperparameters' variation. We also propose to replace the hierarchical gridding method for spatial discretization with a simple linear interpolation technique, which can reflect the continuous spatial distance while providing dense representation. Experimental comparison with baseline models unequivocally demonstrates the superiority of our model, as STAN improves recall rates to new records that surpass the state-of-the-art models by 9-17\%. 

\begin{figure}[t]
\centering
\includegraphics[width=0.85\columnwidth]{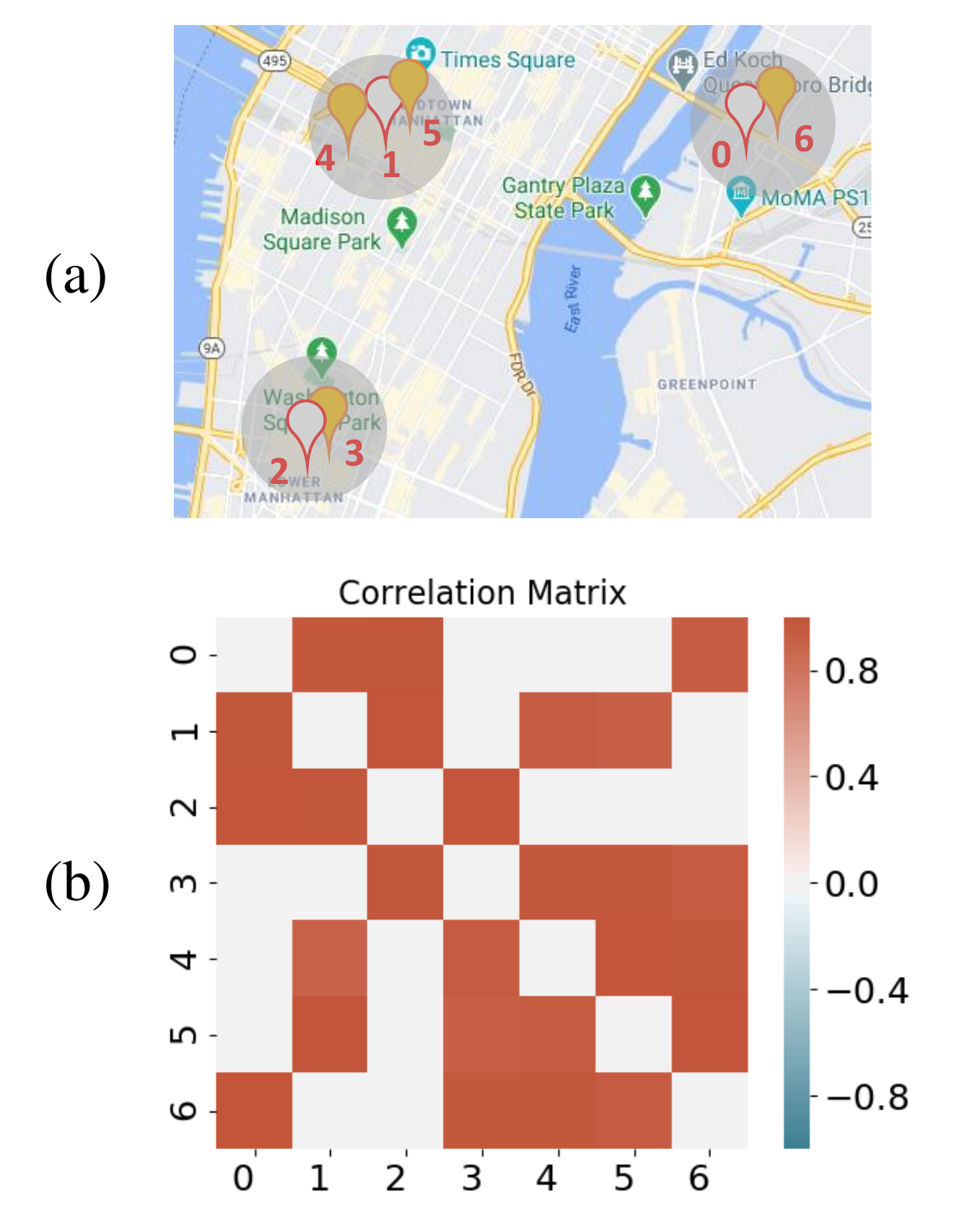}
\caption{The mechanism of STAN. (a) An example map showing the aggregation of visited locations. The locations with the same colored marks and locations within the range of the same dark circles are aggregated. This gives solid evidence that non-adjacent locations may be correlated and aggregated in our model. (b) The correlation matrix. Here, we take the softmax of the multiplication of query and key in the self-attention aggregation layer as a correlation matrix, which is used to update the representation of check-ins.}
\label{fig5}
\end{figure}

%
\begin{acks}
This work is supported by National Key Research and Development Program (2018YFB1402605, 2018YFB1402600), National Natural Science Foundation of China (U19B2038, 61772528), Beijing National Natural Science Foundation (4182066).
\end{acks}

\balance
\bibliographystyle{ACM-Reference-Format}
\bibliography{ref}

\end{document}